\documentclass[aps,pra,twocolumn,showpacs,groupedaddress,superscriptaddress,nofootinbib,floatfix]{revtex4-2}
\usepackage{bm,graphicx}
\usepackage[utf8]{inputenc}
\usepackage[T1]{fontenc}
\usepackage{lmodern}
\usepackage{placeins}
\usepackage{amssymb,amsmath}
\usepackage{epsfig}
\usepackage{natbib}
\usepackage{color}
\usepackage{makecell}
\usepackage{bm}
\usepackage[pdftex,colorlinks,linkcolor=blue,citecolor=blue,urlcolor=blue]{hyperref}
\usepackage{subfigure}
\usepackage{array}
\usepackage{multirow} 
\usepackage{booktabs}
\usepackage{dsfont}
\usepackage[normalem]{ulem}
\newcommand\scalemath[2]{\scalebox{#1}{\mbox{\ensuremath{\displaystyle #2}}}}
\usepackage{color}
\usepackage{ulem}
\usepackage{comment}
\newcommand{\ket}[1]{\ensuremath{\left| #1 \right\rangle}}
\newcommand{\bra}[1]{\ensuremath{\left\langle #1 \right|}}

\newcommand{\ie}{{\it{i.e.,~}}}

\newcommand{\cf}{{\it{cf.~}}}
\newcommand{\refeq}[1]{Eq.~(\ref{#1})}
\newcommand{\refeqs}[1]{Eqs.~(\ref{#1})}
\newcommand{\1}{\ensuremath{\left|1\right\rangle}}
\newcommand{\2}{\ensuremath{\left|2\right\rangle}}
\newcommand{\3}{\ensuremath{\left|3\right\rangle}}
\newcommand{\4}{\ensuremath{\left|4\right\rangle}}

\definecolor{gray}{gray}{0.6}
\begin{document}

\title{Quantum-jump analysis of frequency up-conversion amplification without inversion in a four-level scheme}
\author{J.~L.~Rubio}\email[]{juanluis.rubio@uab.cat}
\affiliation{Departament de F\'{\i}sica, Universitat Aut\`{o}noma de Barcelona, E-08193 Bellaterra, Spain}
\author{J.~Mompart}
\affiliation{Departament de F\'{\i}sica, Universitat Aut\`{o}noma de Barcelona, E-08193 Bellaterra, Spain} 
\author{V.~Ahufinger}
\affiliation{Departament de F\'{\i}sica, Universitat Aut\`{o}noma de Barcelona, E-08193 Bellaterra, Spain}
\date{\today}
\begin{abstract} 
In this work, we use the quantum-jump approach to study 
light-matter interactions in a four-level scheme giving rise to frequency up-conversion amplification without inversion (AWI).
%This approach is based on the Monte Carlo wave-function (MCWF) formalism which provides a description of the light-matter interaction of a single quantum system subjected to dissipative processes. Under some approximations, the QJ approach allows to obtain semi-analytical expressions for the different mechanisms that contribute to the amplification and attenuation of a probe field.
The results obtained apply to the case of neutral Hg vapour where, recently, it has been reported AWI of a probe field in the UV regime, opening the way for lasing without inversion in this range of frequencies. 
We show that, in this scheme, the key element to obtain maximum amplification is the fulfillment of the three-photon resonance condition, which favors coherent evolution periods associated with the probe field gain. We also investigate the parameter values in order to optimize amplification.
The present study extends the theoretical understanding of the underlying mechanisms involved in AWI.
\end{abstract}

%03.67.Dd 	Quantum cryptography and communication security
%03.67.Hk 	Quantum communication
%03.67.Lx 	Quantum computation architectures and implementations
%32.80.Qk 	Coherent control of atomic interactions with photons
%42.50.Md 	Optical transient phenomena: quantum beats, photon echo, free-induction decay, dephasings and revivals, optical nutation, and self-induced transparency
%... ... ... 	Dynamics of nonlinear optical systems; optical instabilities, optical chaos, and optical spatio-temporal dynamics, see 42.65.Sf
%... ... ... 	Optical solitons; nonlinear guided waves, see 42.65.Tg
%42.50.Nn 	Quantum optical phenomena in absorbing, amplifying, dispersive and conducting media; cooperative phenomena in quantum optical systems
%42.50.Gy 	Effects of atomic coherence on propagation, absorption, and amplification of light; electromagnetically induced transparency and absorption
 
\maketitle

\section{INTRODUCTION}
\label{sec:INTRODUCTION}

In the last decades, one of the technical challenges of laser physics has been the development of continuous-wave lasers in the UV and VUV frequency range, with relevant applications in spectroscopy or lithography, among others.
The main constraint is the fact that, in order to obtain population inversion in the laser transition, the threshold pumping power scales with the laser frequency from $\omega^{4}$ to $\omega^{6}$ \cite{Mompart'00}.
A way to circumvent this constraint is to build lasers exploiting nonlinear effects, as using four-wave sum-frequency mixing \cite{Scheid'09, Kolbe'12}.
An alternative is the development of lasers that do not require population inversion, \ie lasing without inversion (LWI) \cite{Scully'89, Kocharovskaia'88}, following the amplification of a probe field without population inversion in the corresponding probe field transition, the so-called amplification without inversion (AWI) \cite{Harris'89, Mompart'00, Nottelmann'93}.
The key idea of AWI is to cancel absorption of a probe field by means of quantum interferences while maintaining unchanged or favouring stimulated emission.
%AWI has been proposed for several atomic three-levels schemes [Ref] and, in particular, in a V-type scheme, which has been reported in cold Rb atoms carried out the basis of the first laser operating with LWI \cite{Zibrov'95}.
However, atomic schemes commonly used in AWI do not allow the wavelength of the probe laser to be significantly smaller than that of the driving lasers used to excite the required coherence, and therefore are far from achieving UV or VUV laser light from an optical driving laser.
Nevertheless, a particular four-level scheme in neutral Hg has been recently investigated experimentally \cite{Rein'15, Rein'22} reporting AWI in the UV frequency range paving the way towards LWI in this spectral domain.

Usually, the schemes involved in the aforementioned works are studied using the density matrix equations (DME), which allow to obtain average values of the magnitudes for an ensemble of atoms. However, this description does not provide detailed information about the different light-matter mechanisms that are present in the interactions.
In contrast to the DME formalism, the Monte Carlo wave-function (MCWF) formalism \cite{Dalibard'92, Dum'92, Plenio'98} enables the study of the time evolution of each individual atomic system, describing its interaction with light as a sequence of coherent evolution periods, \ie time intervals between two consecutive quantum jumps.
This approach, also known as quantum-jump (QJ) or quantum-trajectory method, has been successfully applied to various problems in Quantum Optics \cite{Cohen-Tannoudji'93, Mompart'98,Jong'97} and is closely related with continuous-time measurement schemes \cite{Wiseman'09, Evanseck'00, vanHandel'05, Gross'18, Lewale'20}. MCWF formalism is based on the numerical integration of the time-dependent Schrödinger equation with an effective non-hermitian Hamiltonian and re-normalizing at each time step.
Nevertheless, effective non-hermitian Hamiltonians have been widely used for the study of open systems, especially in optics and electronics, see Ref. \cite{Naghiloo'19, DeCarlo'22, Ashida'20} and references therein.

MCWF formalism allows us to obtain information from the respective contributions of the different dissipative and coherent processes, and the results obtained averaging over many realisations converge with those of the DME \cite{Dum'92}.
Nonetheless, this formalism has two main limitations: (i) it does not offer analytical expressions and (ii) the required computational time is, in general, very long.
The QJ approach \cite{Cohen-Tannoudji'93, Kornyik'19}, based on the MCWF formalism, offers some advantages.
Under certain approximations, it allows us to study the statistical properties of the coherent periods occurring between two successive quantum jumps, as well as to obtain semi-analytical expressions for their occurrence probabilities.
In the case of AWI, the respective contributions of the various physical processes involved in the amplification or attenuation of a probe field can be obtained \cite{Mompart'00}. 
In some schemes, AWI can arise via population inversion in a hidden basis, \ie a meaningful basis, as the so-called CPT \cite{Agape'93} or the dressed-states bases. In other schemes, where AWI occurs without inversion on any hidden basis, the origin of gain can be elucidated by the QJ technique.
In this respect, the use of the QJ approach allows one to explain probe AWI in V- and $\Lambda$-type, which appears at line center, \ie between the two absorption resonances associated with the dressed states, as resulting from the fact that two-photon gain processes overcome two-photon loss process even without two-photon population inversion.
Analogously, for ladder-type schemes, where probe AWI appears at the two sidebands \cite{Mompart'98,Mandel'92} located outside the region delimited by the two dressed-state
resonances, AWI is explained due to the fact that one-photon gain processes overcome one-photon loss processes \cite{Mompart'03}.

The aim of this work is to study, using the QJ approach, a four-level atomic scheme as the one used in \cite{Rein'15} for neutral Hg vapour in order to understand the origin and the conditions to obtain AWI. This approach allows to elucidate the most favourable conditions for the amplification of a probe field and isolates the contributions of each coherent process, opening the way to a better understanding of the experiment.
In Section~\ref{sec:Model6}, we introduce the atomic model under study together with some of the results obtained from the DME with the experimental parameter values used in \cite{Rein'15}.
In Section~\ref{sec:Quantum-jumpanalysis6}, 
a QJ analysis of the scheme is carried out, reviewing, in the first place, 
the main concepts of the MCWF and the QJ approaches.
Then, the most favourable conditions for the probe field amplification are discussed, and the  semi-analytical expressions for the  probabilities  of the coherent evolution processes are derived.
In Sec.~\ref{sec:Numericalresults}, the numerical results
for the optimal parameters and configurations are presented.
Finally, we summarise the results of this work and present the conclusions in Sec.~\ref{sec:Conclusions6}.

\section{Model} %Section - 1.1
\label{sec:Model6}

%
%%%%%%%%%%%%%%%%%%%%%%%%%%%%%%%%%%%%%%%%%%
\begin{figure}[ht!]
\centering
{
\includegraphics[width=0.3\textwidth]{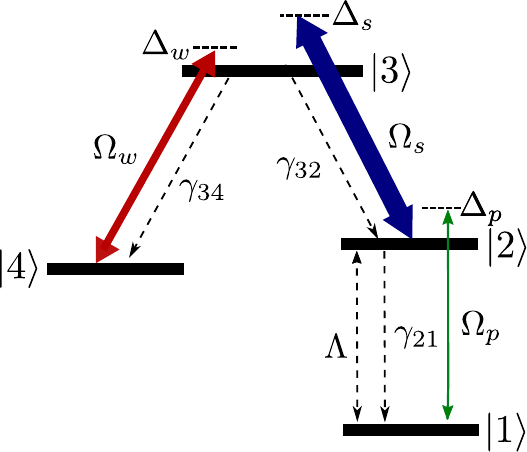}
}
\caption{The four-level atomic scheme under investigation.
$\Omega_{w}$, $\Omega_{s}$, and $\Omega_{p}$ are the Rabi frequencies, and $\Delta_{w}$, $\Delta_{s}$, and $\Delta_{p}$ the  detunings of the weak, strong, and probe laser fields, respectively.
$\gamma_{32}$, $\gamma_{34}$, $\gamma_{21}$ and $\Lambda$ are the rates for the incoherent processes (see the text).
}
\label{fig6.1}
\end{figure}
%%%%%%%%%%%%%%%%%%%%%%%%%%%%%%%%%%%%%%%%%%%%%%
%

Fig.~\ref{fig6.1} shows the four-level scheme under consideration.
In this scheme, a probe laser field couples levels \1$\leftrightarrow$\2 with Rabi frequency $\Omega_{p}$ and detuning $\Delta_{p}$, while a strong (weak) laser field couples levels \3$\leftrightarrow$\2 (\3$\leftrightarrow$\4) with Rabi frequency $\Omega_{s}$ ($\Omega_{w}$) and detuning $\Delta_{s}$ ($\Delta_{w}$).
Population decays by spontaneous emission from level \3 to levels \2 and \4 with rates $\gamma_{32}$ and $\gamma_{34}$, respectively, and from level \2 to level \1 with rate $\gamma_{21}$. In addition, we  consider the presence of a bidirectional incoherent pumping mechanism between levels \1 and \2 with rate $\Lambda$.
We assume that Rabi frequencies are real and, therefore, the sign of ${\rm Im}[\rho_{ij}]$ indicates absorption/amplification of the field coupled to the transition $\ket{i}\leftrightarrow \ket{j}$, being $\rho_{ij}$ the $ij$-element of the density matrix operator.

On the basis of the scheme shown in Fig.~\ref{fig6.1}, we consider the states $\1\equiv\ket{6\;^{1}S_{0}}$, $\2\equiv\ket{6\;^{3}P_{1}}$, $\3\equiv\ket{7\; ^{3}S_{1}}$, and $\4\equiv\ket{6\;^{3}P_{2}}$ of neutral Hg, with decay rates $\gamma_{21}=2\pi\times1.27\,\mu{\rm s}^{-1}$,  $\gamma_{34}=2\pi\times7.75\,\mu{\rm s}^{-1}$, and $\gamma_{32}=2\pi\times8.86\,\mu{\rm s}^{-1}$ (values from the NIST database \cite{NIST}).
This scheme has been recently used \cite{Rein'15} to report AWI of a probe laser field coupled to the transition \1$\leftrightarrow$\2 using a hot vapour of Hg atoms.
%
%%%%%%%%%%%%%%%%%%%%%%%%%%%%%%%%%%%%%%%%%%
\begin{figure}[ht!]
\centering{
\includegraphics[width=0.5\textwidth]{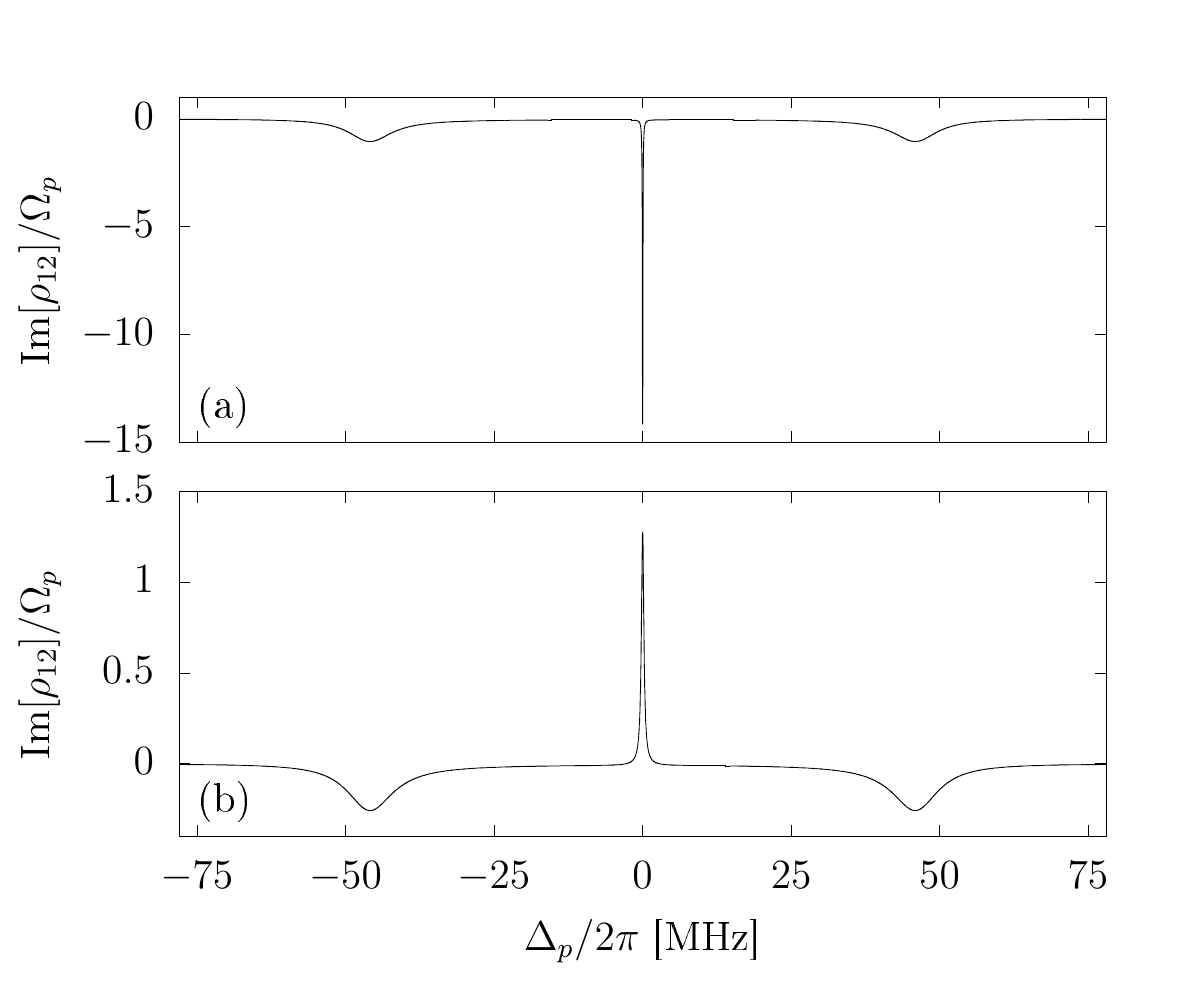}
}
\caption{Im$[\rho_{12}]/\Omega_{p}$ as a function of $\Delta_{p}/2\pi$ without incoherent pumping (a) and
with incoherent pumping rate $\Lambda=2\pi\times0.38\,\mu{\rm s}^{-1}$ (b). The parameter values are $\Omega_{s}=2\pi\times88.9$ MHz, $\Omega_{w}=2\pi\times25.4$ MHz, $\Omega_{p}=2\pi\times0.0013$ MHz, and  $\Delta_s=\Delta_w=0$. See the text for the rest of parameter values corresponding to Hg atomic transitions. Im$[\rho_{12}]/\Omega_{p}<0\;(>0)$ indicates absorption (amplification) of the probe field.}
\label{fig6.9}
\end{figure}
%%%%%%%%%%%%%%%%%%%%%%%%%%%%%%%%%%%%%%%%%%%%%%
%
%Note that the parameter values for this scheme are quite similar to those considered in the previous sections.
%Therefore, taking into account our previous discussions, 
%By means of simulations based on DME formalism, a set of optimum values have been established to obtain a probe field amplification peak, obtaining similar to those reported in the experiment.

Fig.~\ref{fig6.9} shows Im$[\rho_{12}]/\Omega_{p}$ as a function of $\Delta_{p}$  obtained by solving the DME of the system using $\Delta_{s}=\Delta_{w}=0$ without an incoherent pumping rate (a)
and with an incoherent pumping rate
$\Lambda=2\pi\times0.38\,\mu{\rm s}^{-1}$  (b).
The Rabi frequency values have been set to:
$\Omega_{s}=2\pi\times88.9$ MHz, $\Omega_{w}=2\pi\times25.4$ MHz, and $\Omega_{p}=2\pi\times0.0013$ MHz.
Note that a single parameter, the pumping rate transforms the response of the medium to the probe field from absorption (a) to amplification (b) at $\Delta_p=0$.
Note also that amplification takes place at $\Delta_p=\Delta_w=\Delta_w=0$ so one-photon, two-photon, and three-photon resonance conditions are satisfied. As we will see later on, amplification is due to the dominant role of three-photon gain processes. These three-photon processes are Doppler-free
%Note how the optimal value for the pumping rate inverts the behaviour of the probe field in the absence of pumping by converting an absorption peak (${\rm Im[}\rho_{12}]<0$) into an amplification peak (${\rm Im[}\rho_{12}]>0$) for $\Delta_{p}=0$.
%We can see that for a three-photon resonance configuration, \ie $\Delta_{3}\equiv\Delta_{p}+\Delta_{s}-\Delta_{w}=0$, trivially satisfied in the experiment with $\Delta_{p}=\Delta_{s}=\Delta_{w}=0$, we have a narrow peak of probe amplification for a set of optimal values.
%---------------------------------------------------------
%In this case, it can be expressed the condition as $\Delta_{3}\equiv\Delta_{p}+\Delta_{s}-\Delta_{w}=0$, trivially satisfied if $\Delta_{p}=\Delta_{s}=\Delta_{w}=0$.
%Indeed, let us suppose that the above condition is fulfilled for atoms at rest.
%Due to the Doppler effect, an atom with a velocity $\overrightarrow{v}$ in interaction with an  electromagnetic field with nominal frequency $\omega$ and wave vector $\overrightarrow{k}$ sees a Doppler shifted frequency $\omega^{\prime}=\omega-\overrightarrow{k}\cdot\overrightarrow{v}$.
%In this case, the three-photon resonance condition is modified as
%
%\begin{equation}
%\label{eq:Delta3}
%\Delta_{3}\equiv\Delta_{p}+\Delta_{s}-\Delta_{w}-(\overrightarrow{k_{p}}+\overrightarrow{k_{s}}-\overrightarrow{k_{w}})\cdot\overrightarrow{v}=0.
%\end{equation}
%
provided that the interacting electromagnetic fields are properly oriented, specifically, that their wave-vectors satisfy $\overrightarrow{k_{p}}+\overrightarrow{k_{s}}-\overrightarrow{k_{w}}=0$ \cite{Rein'15}.

%Considering the wave vectors of the interacting electromagnetic fields, $k_{i}$, it can be shown that if an atom at rest fulfils the three-photon resonance condition, all the moving atoms will also fulfil it provided that $\overrightarrow{k_{p}}+\overrightarrow{k_{s}}-\overrightarrow{k_{w}}=0$, \ie through a proper orientation of the fields wave vectors, as used in Ref.~\cite{Rein'15}. Therefore, in hot vapours with large inhomogeneous broadening all atomic velocity classes can contribute to AWI in this Doppler-free configuration.

\begin{figure}[ht!]
    \includegraphics[width=0.5\textwidth]{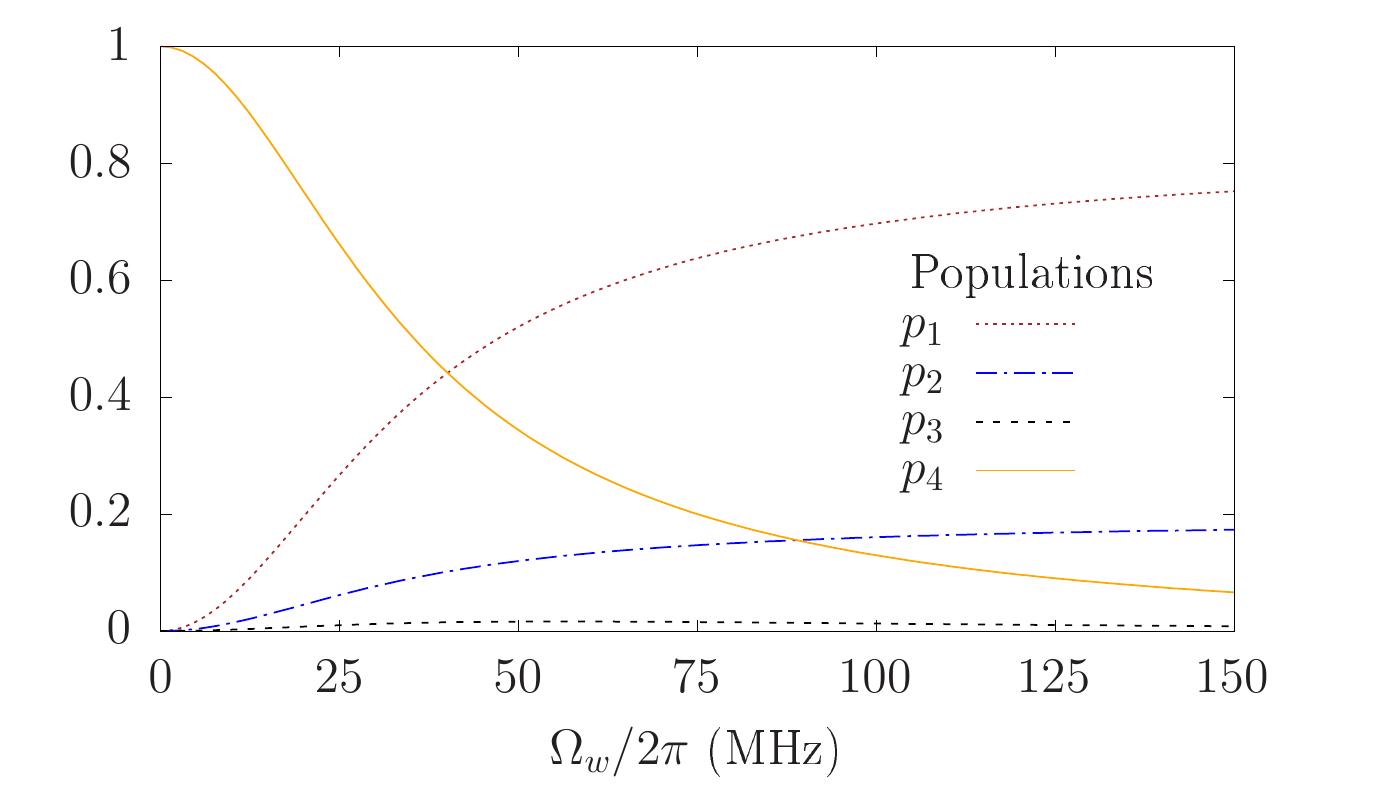}
    \caption{Steady-state populations of the atomic levels \ket{1} ($p_{1}$), \ket{2} ($p_{2}$), \ket{3} ($p_{3}$), and \ket{4} ($p_{4}$) of the scheme under investigation as a function of $\Omega_{w}$.
    %(b) Im$[\rho_{12}]/\Omega_{p}$ as a function of $\Omega_{s}$ and $\Omega_{w}$.
    The rest of parameter values are the same as in Fig.~\ref{fig6.9}(b).}
    \label{figPaper1}
\end{figure}

Complementing our previous analysis by using the DME approach, Fig.~\ref{figPaper1} shows the steady-state populations 
of the four atomic states as a function of $\Omega_{w}$, with the rest of parameter values as in Fig.~\ref{fig6.9}(b). We observe that there is no population inversion between transitions \1$\leftrightarrow$\2 and \1$\leftrightarrow$\3 for any value of $\Omega_{w}$, and there is also no population inversion in the transition \1$\leftrightarrow$\4 for $\Omega_{w}/2\pi\geq40$ MHz.
%Fig.~\ref{figPaper1}(b) shows Im$[\rho_{12}]/\Omega_{p}$ as a function of $\Omega_{s}$ and $\Omega_{w}$. We observe that, by fixing $\Omega_{s}\sim2\pi\times90$ MHz, AWI is obtained for a range of values $\Omega_{w}=0-50$ MHz, with a maximum for $\Omega_{w}\sim2\pi\times25$ MHz. It is also observed that the maximum amplification of the probe field is obtained for a ratio between the weak and the strong fields Rabi frequencies of $\Omega_{w}/\Omega_{s}\sim0.2$.

In the following, a study of this scheme using the QJ approach will be carried out to elucidate the physical processes responsible for AWI in this four level scheme and derive the optimal values that maximize gain.

%
%%%%%%%%%%%%%%%%%%%%%%%%%%%%%%%%%%%%%%%%%%
%%%%%%%%%%%%%%%%%%%%%%%%%%%%%%%%%%%%%%%%%%%%%%
%
%
%%%%%%%%%%%%%%%%%%%%%%%%%%%%%%%%%%%%%%%%%%
\begin{figure*}[ht!]
\centering
{
\includegraphics[width=0.9\textwidth]{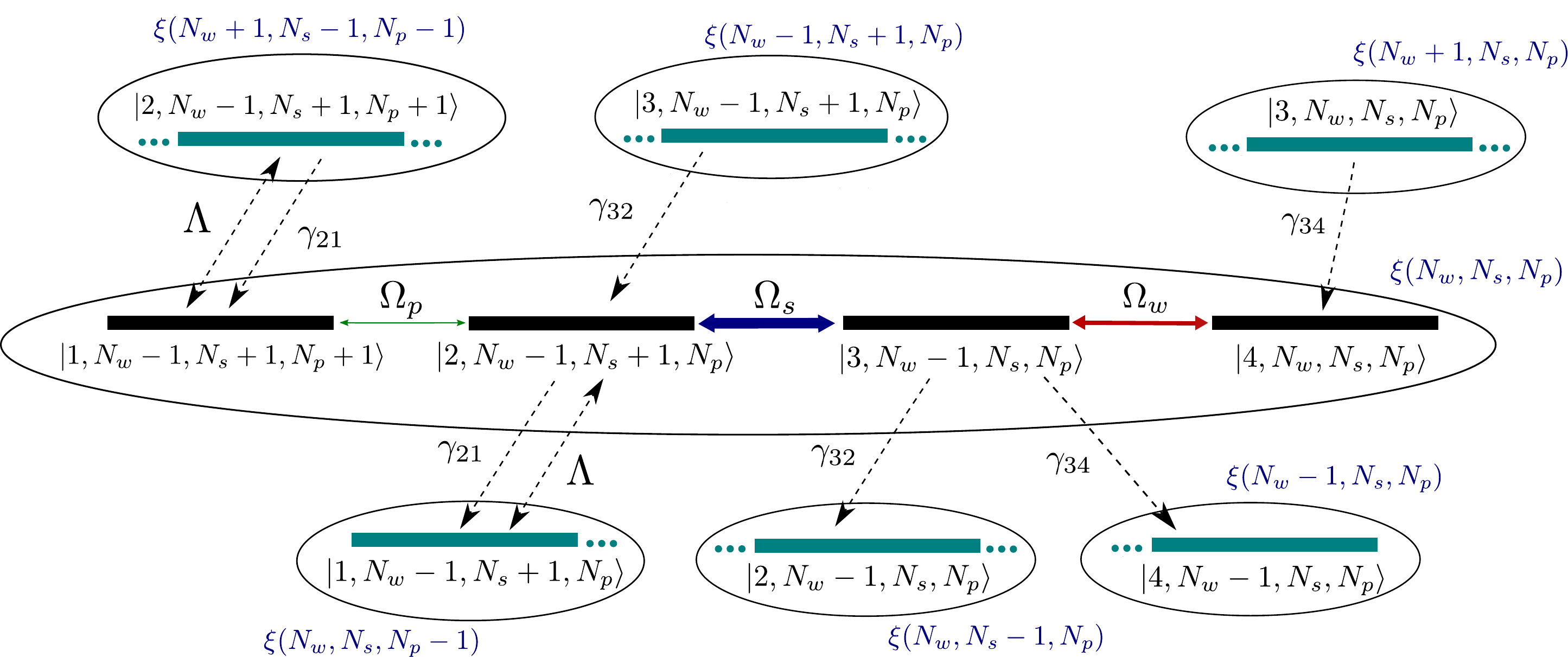}
}
\caption{Manifolds of the four states of the atom+laser photon numbers. $N_{w}$, $N_{s}$ and $N_{p}$ are the photon numbers of the weak, strong and probe laser field, respectively. The continuous evolution given by the coherent interaction is shown with solid arrows and the dissipative processes are shown with dashed arrows, accounting for the quantum jumps associated to spontaneous emission and incoherent pumping.}
\label{fig6.2}
\end{figure*}
%%%%%%%%%%%%%%%%%%%%%%%%%%%%%%%%%%%%%%%%%%%%%%
%

\section{Quantum-jump analysis} %Section - 1.1
\label{sec:Quantum-jumpanalysis6}

%In this section, the main concepts of the QJ approach will be reviewed and applied to the considered four-level scheme.
From Fig.~\ref{fig6.2} one can identify the different coherent periods and quantum jumps that can occur at random times on the scheme under consideration (see Fig.~\ref{fig6.1}). Using combined atom+field states, each manifold of four states is labelled as
\begin{align}\label{eq:manifold}
&\xi(N_{w}+m,N_{s}+n,N_{p}+q)\equiv \nonumber \\
&\{\ket{1,N_{w}+m-1,N_{s}+n+1,N_{p}+q+1}, \nonumber \\
&\ket{2,N_{w}+m-1,N_{s}+n+1,N_{p}+q}, \nonumber \\
&\ket{3,N_{w}+m-1,N_{s}+n,N_{p}+q}, \nonumber \\
&\ket{4,N_{w}+m,N_{s}+n,N_{p}+q}\},
\end{align}
being $m,n,q = 0,\pm1,\pm2,\dots$, and $N_{p}$, $N_{s}$ and $N_{w}$ the number of photons in the probe, strong, and weak laser fields, respectively.
The solid arrows (in colour) represent the interaction of a laser with a pair of atomic levels in a specific manifold and the dashed arrows (in black) represent the dissipative processes, \ie pumping and spontaneous emission, between two different manifolds.
This description enables the visualization of the coherent processes within a manifold and incoherent ones between manifolds for a single atom, so that it allows us to count the photons emitted/absorbed in each of the fields for a single atom.
We consider that all fields are intense enough such the photon number fluctuations are negligible compared to their corresponding mean photon number. Thus, $N_{p}$, $N_{s}$, and $N_{w}$ can be  interpreted as the mean photon number of the probe, strong, and weak fields, respectively.

The interpretation of the diagram shown in Fig.~\ref{fig6.2} is as follows.
One quantum jump places the state of the system in the atomic state \ket{i}, with $i=1,2,3,4$, of one of the manifolds. 
Then, the system evolves within that manifold, following a coherent period that begins in the initial state \ket{i}, where the previous quantum jump ended, and that ends in the state \ket{j} of the manifold from which the next quantum jump occurs.
We will denote such a coherent period as period$(i,j)$.
On the other hand, the dissipative processes responsible for the quantum jumps are denoted by $J_{ij}$,
meaning that the quantum jump occurs from state \ket{i} to state \ket{j}.
Thus, the evolution of the system is described by a quantum trajectory, which is a sequence with all the coherent periods and the quantum jumps between them, 
\begin{equation}
\dots J_{ij}\,{\rm period}(j,k)\,J_{kl}\,{\rm period}(l,r)\,J_{rs}\dots\nonumber
\end{equation}
where \ket{i},\ket{j},\ket{k},\ket{l},\ket{r}, and \ket{s} are atomic states.
Each of the coherent periods implies an increase or decrease in the photon number of the fields involved.

In the QJ approach, the evolution of the wave-function is given by the Schrödinger equation
\begin{align}\label{eq:Schrödinger9}
i\hslash\frac{d}{dt}\ket{\psi(t)}=\hat{H}_{{\rm NH}}\ket{\psi(t)},
\end{align}
where
\begin{equation} \label{eq:Hnh}
{\hat{H}_{\rm NH}}=-\cfrac{\hslash}{2}\left(
\scalemath{0.85}{
\begin{array}{cccc}
    iG_{1} & \Omega_{p} & 0 & 0 \\
    \Omega_{p} & -2\Delta_{p}+iG_{2} & \Omega_{s} & 0 \\
		0 & \Omega_{s} & -2\Delta_{s}+iG_{3} & \Omega_{w} \\
		0 & 0 & \Omega_{w} & -2(\Delta_{s}-\Delta_{w})
  \end{array}} 
  \right)
\end{equation}
is the non-hermitian Hamiltonian under the electric-dipole (EDA),  and rotating-wave (RWA) approximations and in the interaction picture. $G_{i}$ is the total departure rate due to incoherent processes from the atomic state \ket{i}, being $G_{1}=\Lambda$, $G_{2}=\gamma_{21}+\Lambda$, and $G_{3}=\gamma_{34}+\gamma_{32}$.
Note that the non-hermitian Hamiltonian can be written as
\begin{equation}\label{eq:Hnh1}
\hat{H}_{\rm NH}=\hat{H}_{0}-\cfrac{i\hslash}{2}\left(G_{1}\hat{S}_{21}\hat{S}^{\dagger}_{21}+G_{2}\hat{S}_{32}\hat{S}_{32}^{\dagger}+G_{3}\hat{S}_{34}^{\dagger}\hat{S}_{34}\right),
\end{equation}
with
\begin{align}\label{eq:Hnh1}
\hat{H}_{0}=\hslash\left(\Delta_{p}\hat{S}_{21}^{\dagger}\hat{S}_{21}+\Delta_{s}\hat{S}_{32}^{\dagger}\hat{S}_{32}+(\Delta_{s}-\Delta_{w})\hat{S}_{34}\hat{S}_{34}^{\dagger}\right) \nonumber\\
-\cfrac{\hslash}{2}\left(\Omega_{p}(\hat{S}_{21}^{\dagger}+\hat{S}_{21})+
\Omega_{s}(\hat{S}_{32}^{\dagger}+\hat{S}_{32})+
\Omega_{w}(\hat{S}_{34}^{\dagger}+\hat{S}_{34})\right)
\end{align}
the hermitic Hamiltonian in the interaction picture assuming EDA and RWA, where $\hat{S}_{ij}=\ket{j}\bra{i}$, and $\hat{S}_{ij}^{\dagger}=\ket{i}\bra{j}$ are the lowering and raising atomic operators, respectively, associated with electric-dipole allowed atomic transitions.

We can obtain a quantum trajectory of the system using the MCWF formalism. 
In this approach, the wave-function is evolved for very small intervals of time $dt\ll G_{i}^{-1}$ using the non-unitary evolution operator built with the non-hermitian Hamiltonian.
This temporal evolution is conditioned by imaginary (\textit{Gedankenexperiment}) and random detections of the collapses due to the dissipative processes.
In each $dt$, the wave-function can either undergo a quantum jump and collapse into a state or continue in a superposition following a coherent evolution after which it must be normalised, because the no-jump evolution described by \refeq{eq:Schrödinger9} and \refeq{eq:Hnh} does not occur with probability
one.
As mentioned above, the average of these quantum trajectories over an atomic ensemble reproduces the DME results. These are derived from the Lindblad master equation, which describes the temporal evolution of the density matrix $\hat{\rho}$, 
% The temporal evolution of density matrix, $\hat{\rho}$, is given by the Schrödinger--von Neumann equation. The terms accounting for the incoherent or dissipative processes are added \textit{ad hoc} in the compact form $\hat{L}\hat{\rho}$, resulting
% %
\begin{align}\label{eq:Schrödinger-vonNeuman-Liouville}
\dot{\hat{\rho}}(t)=-\frac{i}{\hslash}[\hat{H_{0}}(t),\hat{\rho}(t)]+\hat{L}\hat{\rho}(t),
\end{align}
being, in our case,
% where $\hat{L}$ is the Liouville operator and \refeq{eq:Schrödinger-vonNeuman-Liouville} the Schrödinger--von Neumann--Liouville equation.
% The system of equations obtained from \refeq{eq:Schrödinger-vonNeuman-Liouville} are the so-called density matrix equations (DME).
% For spontaneous emission, the \textit{ad hoc} term is defined as a Lindblad operator taking the form
%
\begin{align}\label{eq:Liouville}
%\label{eq:Lrho}
\hat{L}\hat{\rho}(t)=&\sum_{i\neq j}\cfrac{\gamma_{ij}}{2}\left(2\hat{S}_{ij}\hat{\rho}(t)\hat{S}_{ij}^{\dagger}-\hat{S}_{ij}^{\dagger}\hat{S}_{ij}\hat{\rho}(t)-\hat{\rho}(t)\hat{S}_{ij}^{\dagger}\hat{S}_{ij}\right),
\end{align}
the Lindblad operator. 
In the last expression, $\ket{i}\rightarrow\ket{j}$ is an allowed transition, $\gamma_{ij}$ is its corresponding decay rate (including spontaneous decay  plus incoherent pumping, if applicable). In our scheme the electric-dipole allowed transitions are $\ket{1}\rightarrow\ket{2}$, $\ket{2}\rightarrow\ket{1}$, $\ket{3}\rightarrow\ket{2}$, and $\ket{3}\rightarrow\ket{4}$, and
the correspondence between the quantum jumps $J_{ij}$ with the lowering and raising atomic operators is $J_{12}\leftrightarrow\hat{S}_{21}^{\dagger}$, $J_{21}\leftrightarrow\hat{S}_{21}$, $J_{32}\leftrightarrow\hat{S}_{32}$, and $J_{34}\leftrightarrow\hat{S}_{34}$. Note that there are no quantum jumps ending in \ket{3} or starting from \ket{4}.
% The application of \refeq{eq:Liouville} is equivalent to add the following terms in the DME: $(\dot{p_{i}})_{{\rm incoh}}\propto+\sum_{E_{j}>E_{i}}\gamma_{ji}p_{j}-\sum_{E_{i}>E_{j}}\gamma_{ij}p_{i}$ in the equations for the populations, and $(\dot{\rho_{ij}})_{{\rm incoh}}\propto-\Gamma_{ij}\rho_{ij}$, where $\Gamma_{ij}=(1/2)(\gamma_{i}+\gamma_{j})$, being $\gamma_{i}$ ($\gamma_{j}$) the total population departure rate from state \ket{i} (\ket{j}), in the equations for the coherences.

Using the MCWF formalism, numerical values for the probabilities of each coherent period between two quantum jumps can be obtained.
However, it is also possible to obtain semi-analytical expressions in certain particular cases without the need of numerical simulations, as we will show in the following.
%This approach allows an analysis of the most favourable conditions for achieving amplification of the probe field, as will be discussed in the following Section.

We are interested in studying the amplification of the probe field in the four-level scheme under consideration (see Fig.~\ref{fig6.1}).
Note that, due to the absence of dissipative processes that collapse the wave-function in state \3, there are no coherent periods which start in this state.
In addition, as there are no dissipative processes from state \4, no coherent periods will end in that state.
Therefore, there are 9 different possible coherent periods that can occur in the quantum trajectory of the system: period$(1,1)$, period$(1,2)$, period$(1,3)$, period$(2,1)$, period$(2,2)$, period$(2,3)$, period$(4,1)$, period$(4,2)$, and period$(4,3)$.
However, only four of them involve changes in the photon number of the probe laser field, $\Delta N_{p}$.
Table~\ref{table6.1} lists those periods indicating, in each case, the respective photon variation for the probe, strong and weak laser fields, the type of process and the net effect, \ie whether it implies gain or loss for the probe field.
%
%%%%%%%%%%%%%%%%%%%%%%%%%%%%%%%%%%%%%%%%%%%
\begin{table}[ht!]
\centering
\scalemath{0.95}{
%\begin{tabular}{@{}|c|c|c|c|c|c|@{}}
\begin{tabular}{|l|c|c|c|c|c|c|l|}
%\toprule
\hline
 & $\Delta N_{p}$ & $\Delta N_{s}$ & $\Delta N_{w}$ & type & effect \\ \hline
\rm{Period}(2,1) & 1 & 0 & 0 & \rm{One\--photon} & \rm{Gain} \\ \hline
\rm{Period}(1,2) & -1 & 0 & 0 & \rm{One\--photon} & \rm{Loss} \\ \hline
\rm{Period}(4,1) & 1 & 1 & -1 & \rm{Three\--photon} & \rm{Gain} \\ \hline
\rm{Period}(1,3) & -1 & -1 & 0 & \rm{Two\--photon} & \rm{Loss} \\
\hline
%\bottomrule
\end{tabular}
}
\caption{Coherent periods that involve changes in the photon number of the
probe laser field. From left to right, the columns indicate the photon number variation of the probe, strong, and weak laser field, the process type and whether the process involves gain or loss of the probe laser field (effect).}
\label{table6.1}
\end{table}
%%%%%%%%%%%%%%%%%%%%%%%%%%%%%%%%%%%%%%%%%%%
%
Therefore, the total mean variation of the probe photon number can be expressed as
\begin{equation}\label{eq:meanchangeprove}
\langle\Delta N_{p}\rangle_{T}=P(2,1)+P(4,1)-P(1,2)-P(1,3),
\end{equation}
where $P(i,j)$ is the probability that a random choice among all coherent evolution periods of the stochastic quantum trajectory results in the period$(i,j)$. 
This probability can be expressed \cite{Cohen-Tannoudji'93} as
\begin{equation}\label{eq:Pij}
P(i,j)=P(i)\,G_{j}\int^{\infty}_{0} |c_{ij}(\tau)|^{2}d\tau,
\end{equation}
where $P(i)$ is the probability that a coherent evolution starts in state $\ket{i}$, and $c_{ij}(\tau)=\bra{j}{\rm exp}(-i\hat{H}_{nh}\tau/\hbar)\ket{i}$ is the probability amplitude of finding the system in state $\ket{j}$ at $t+\tau$ once the coherent evolution period has started at time $t$ in state $\ket{i}$ of the same manifold.
In some limits, we can obtain analytically the probabilities $P(i)$ by means of the recursive relation
\begin{equation}\label{eq:Pi}
P(i)=\sum_{j}P(j)Q(i/j),
\end{equation}
where $Q(i/j)$ is the conditional probability of starting a coherent period in state $\ket{i}$ when the previous one started in state $\ket{j}$.

To obtain the expressions of $Q(i/j)$ we will consider the limit $\Omega_{p}\ll\Lambda,\gamma_{21}$, and $\Omega_{s},\Omega_{w}>\gamma_{34},\gamma_{32},\Lambda$, having in mind the scheme of Fig.~\ref{fig6.1}.
First, we find the values for $Q(i/1)$. Consider that the system started a coherent period in the state \ket{j}=\1.
Since $\Omega_{p}\ll\Lambda,\gamma_{21}$, before the probability of being in any of the other states of the manifold is significant, it is very likely that a quantum jump occurs from the state \1. This quantum jump can only take place via $\Lambda$ and collapses the wave-function in \2, so the next coherent period will begin in this state. Therefore,
\begin{subequations}\label{eq:Qj1}
\begin{align}
Q(1/1)&=0, \\
Q(2/1)&=1, \\
Q(3/1)&=0, \\
Q(4/1)&=0.
\end{align}
\end{subequations}

Let us now consider that a coherent period begins in any of the states \ket{j}=\2,\3,\4. 
In that case, since $\Omega_{s},\Omega_{w}>\gamma_{34},\gamma_{32},\Lambda$, the system evolves into a superposition of all those states while the  probability amplitude of finding the system in \1 will be negligible.
Therefore, the conditional probabilities will be the same regardless of where the previous coherent period began, \ie $Q(i/2)=Q(i/3)=Q(i/4)$, for all \ket{i}.
The probability that the next coherent period starts in a state \ket{i} will be given by the ratio between the number of favourable cases, \ie those that take  the system through a sequence period$(j,l)\,J_{li}$ period$(i,k)$, where \ket{l} and \ket{k} indicate any atomic state, and the number of all the possible cases, \ie those which first cause a period$(j,l)$, then a quantum jump, and then any other coherent period.
This ratio corresponds to the ratio between the sum of the dissipative processes rates which cause the system to collapse in \ket{i} and the total sum of the departure rates of the states \ket{j}=\2,\3, and \4.
Note that $Q(3/j)=0$ for all \ket{j} since it is not possible for the next coherent period to start in \3, because no dissipative process has previously collapsed the wave-function to that state. 
Therefore,
\begin{subequations}\label{eq:Qj2}
\begin{align}
Q(1/2)&=Q(1/3)=Q(1/4)=\cfrac{\gamma_{21}+\Lambda}{D}, \\
Q(2/2)&=Q(2/3)=Q(2/4)=\cfrac{\gamma_{32}}{D}, \\
Q(3/2)&=Q(3/3)=Q(3/4)=0, \\
Q(4/2)&=Q(4/3)=Q(4/4)=\cfrac{\gamma_{34}}{D},
\end{align}
\end{subequations}
where $D=\gamma_{34}+\gamma_{32}+\gamma_{21}+\Lambda$.
By introducing \eqref{eq:Qj1} and \eqref{eq:Qj2} into \eqref{eq:Pi} and \eqref{eq:Pij}, we obtain
\begin{subequations}\label{eq:Pijs}
\begin{align}
\label{eq:Pijsa}
P(2,1)&=\cfrac{\gamma_{21}+\gamma_{32}+\Lambda}{D^{\prime}}\,\Lambda\int^{\infty}_{0} |c_{12}(\tau)|^{2}d\tau, \\
\label{eq:Pijsb}
P(1,2)&=\cfrac{\gamma_{21}+\Lambda}{D^{\prime}}\,(\gamma_{21}+\Lambda)\int^{\infty}_{0} |c_{12}(\tau)|^{2}d\tau, \\
\label{eq:Pijsc}
P(4,1)&=\cfrac{\gamma_{34}}{D^{\prime}}\,\Lambda\int^{\infty}_{0} |c_{41}(\tau)|^{2}d\tau, \\
\label{eq:Pijsd}
P(1,3)&=\cfrac{\gamma_{21}+\Lambda}{D^{\prime}}\,(\gamma_{34}+\gamma_{32})\int^{\infty}_{0} |c_{13}(\tau)|^{2}d\tau,
\end{align}
\end{subequations}
where $D^{\prime}=\gamma_{34}+\gamma_{32}+2(\gamma_{21}+\Lambda)$, and where it has been used that $c_{12}=c_{21}^{*}$ \cite{Cohen-Tannoudji'93}.
%By inspecting the above expressions we can obtain favourable conditions for the amplification of the prove laser field.
Note that all expressions consist of a multiplicative factor that explicitly depends on the pumping and decay rates and the integrals of the modulus square of the probability amplitudes, $\int^{\infty}_{0} |c_{ij}(\tau)|^{2}d\tau$.
These integrals are maximised (minimised) when the system is on (out of) resonance with the corresponding transition \ket{i}$\leftrightarrow$\ket{j}.

\subsection{Gain conditions per process type}

From each type of process (see the previous discussion and Table \ref{table6.1}) we enumerate the conditions to obtain amplification for the probe laser field.

\subsubsection{One-photon processes} 
From expressions \eqref{eq:Pijsa} and \eqref{eq:Pijsb}, we obtain a condition for which the gain exceeds the loss for the probe laser field due to one-photon processes, \ie $P(2,1)>P(1,2)$, 
\begin{equation}\label{eq:lambdacond1}
\Lambda>\cfrac{\gamma^{2}_{21}}{\gamma_{32}-\gamma_{21}},
\end{equation}
valid as long as $\gamma_{32}>\gamma_{21}$. 
Note that for $\gamma_{21}\geq\gamma_{32}$ it is no possible to obtain amplification of the probe field
exclusively with one-photon processes.

\subsubsection{Two-photon processes} 
As $P(3,1)=0$, we only have losses due to two-photon processes, with an occurrence probability given by $P(1,3)$. In order to minimise $P(1,3)$,
it is convenient, in general, to be out of two-photon resonance.

\subsubsection{Three-photon processes}
We have probe gain via the process $P(4,1)$, which can be maximised if the system is under the three-photon resonance condition, \ie $\Delta_{w}=\Delta_{s}+\Delta_{p}$. There are no three-photon loss processes, \ie $P(1,4)=0$.
% %
% %%%%%%%%%%%%%%%%%%%%%%%%%%%%%%%%%%%%%%%%%%
% \begin{figure*}[ht!]
% \centering
% {
% \includegraphics[width=0.8\textwidth]{fig5new.pdf}
% }
% \caption{(a) Total mean variation of the probe photon number per period with respect to the probe detuning $\Delta_{p}$, using $\Delta_{s}=0$, and $\Delta_{w}=25\gamma_{21}$, which exhibits an amplification peak around $\Delta_{p}=\Delta_{w}$ and two wide absorption peaks around $\Delta_{p}=\pm\Omega_{s}/2$. (b-c) Curves for the involved $P(i,j)$ of the corresponding regions of interest. (c) When the three-photon resonance condition is fulfilled, \ie $\Delta_{p}=\Delta_{w}$, the maximum of $P(4,1)$ induces, in turn, a lower maximum of $P(1,3)$. Rest of parameter values as in Fig.~\ref{fig6.3} (d).}
% \label{fig6.5}
% \end{figure*}
% %%%%%%%%%%%%%%%%%%%%%%%%%%%%%%%%%%%%%%%%%%%%%%
% %

In conclusion, the most suitable conditions for obtaining AWI of the probe laser field will be given by the fulfilment of the three-photon resonance condition avoiding the two-photon resonance condition.

\section{Numerical results}
\label{sec:Numericalresults}
\subsection{Favourable configuration}

\begin{figure}[ht!]
    \centering
    \subfigure{\includegraphics[width=0.68\linewidth]{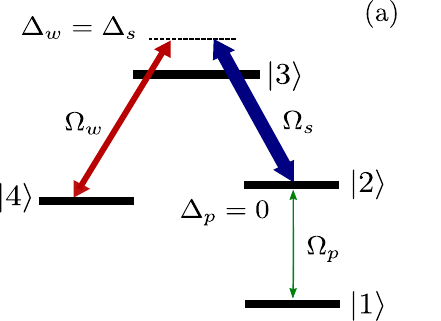}}\\
    \subfigure{\includegraphics[width=1\linewidth]{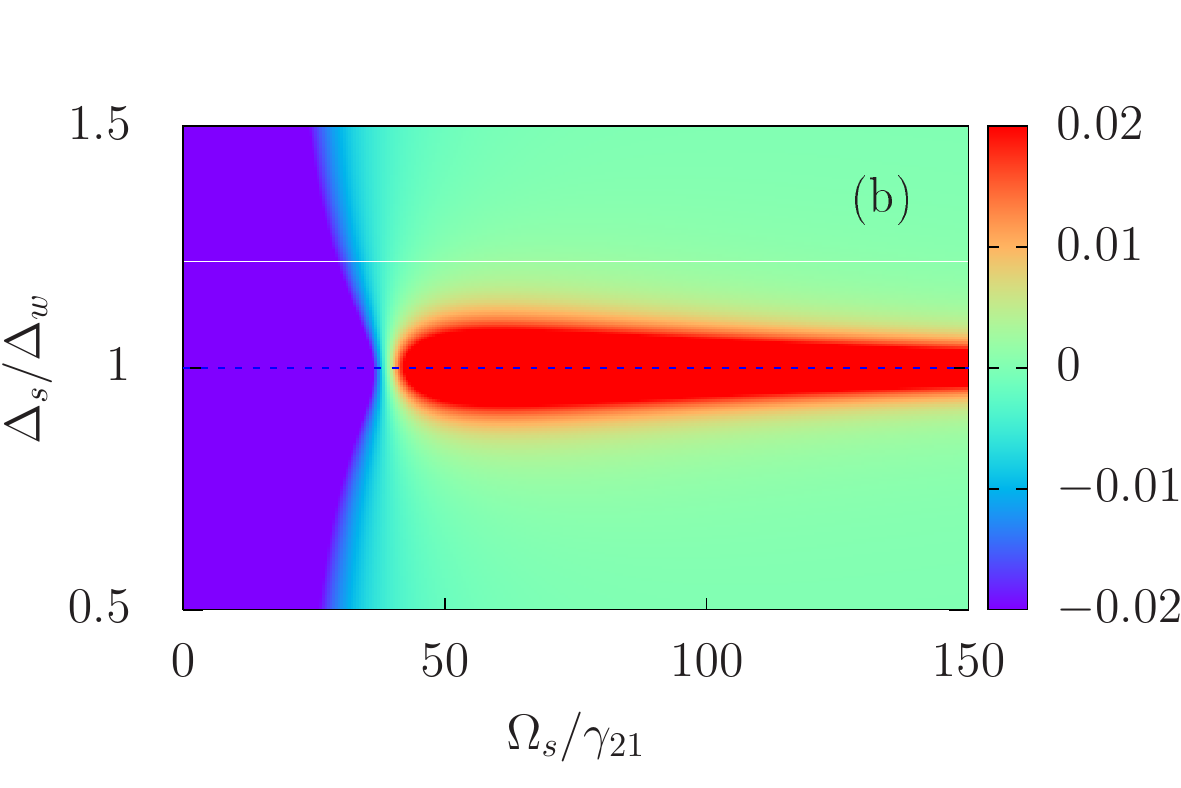}}
    \caption{(a) The most favourable configuration for AWI of the probe laser field in which the probe laser field is on resonance and the system remains under the three-photon resonance condition avoiding the two-photon resonance with transition \ket{1}$\leftrightarrow$\ket{3} if $\Delta_{w}\neq0$.
    (b) ${\rm Im}\left[\rho_{12}\right]/\Omega_{p}$ as a function of $\Omega_{s}/\gamma_{21}$ and $\Delta_{s}/\Delta_{w}$.
    We consider the configuration of (a) using $\Delta_{w}=5\gamma_{21}$. The rest of parameter values are: $\gamma_{32}=5\gamma_{21}$, $\gamma_{34}=10\gamma_{21}$, $\Omega_{w}=20\gamma_{21}$, $\Lambda=0.3\gamma_{21}$, $\Omega_{p}=0.001\gamma_{21}$ (see text for details).}
    \label{fig6.3}
\end{figure}

The most favorable configuration for obtaining AWI in the probe field is represented in Fig.~\ref{fig6.3}(a), in which the probe field is kept at resonance. In this configuration, the three-photon resonance condition is guaranteed, and, provided that $\Delta_{w}\neq0$, the two-photon resonance is avoided at the transition \ket{1}$\leftrightarrow$\ket{3}.
For clarity, the dissipative processes have been omitted in the diagram.
In this scheme, $P(4,1)$ is the relevant process that contributes to the probe gain.
An alternative three-photon resonance configuration is possible (not depicted), in which it is the strong field that is kept in resonance and the probe field is coupled with a certain detuning. However, in this case, the AC-Stark splitting of level \2 produced by the strong field favors the absorption of the probe field near $\Delta_{p}\sim\pm\Omega_{s}/2$ values, competing with the coherent processes that give rise to probe field amplification.

Fig.~\ref{fig6.3}(b) shows ${\rm Im}\left[\rho_{12}\right]/\Omega_{p}$ as a function of $\Omega_{s}/\gamma_{21}$ and $\Delta_{s}/\Delta_{w}$, using $\Delta_{w}=5\gamma_{21}$ and $\Delta_{p}=0$, for the scheme of Fig.~\ref{fig6.3}(a), being $\rho_{12}$ the coherence between levels \1$\leftrightarrow$\2, which indicates probe absorption (amplification) if it is negative (positive). The parameter values are $\gamma_{32}=5\gamma_{21}$, $\gamma_{34}=10\gamma_{21}$, $\Lambda=0.3\gamma_{21}$, $\Omega_{p}=0.001\gamma_{21}$, $\Omega_{w}=20\gamma_{21}$.
The maximum amplification is observed for $\Omega_{s}$ above a certain value $\Omega_{s}\sim 40\gamma_{21}$ when the three-photon resonance condition is fulfilled, \ie around $\Delta_{s}=\Delta_{w}$.

%%%%%%%%%%%%%%%%%%%%%%%%%%%%%%%%%%%%%%%%%%%%%%%%%%%%%%%%%%%%%%%%%%%%%%%%%%%%%
%

%
%%%%%%%%%%%%%%%%%%%%%%%%%%%%%%%%%%%%%%%%%%
\begin{figure*}[ht!]
\centering
{
\includegraphics[width=0.9\textwidth]{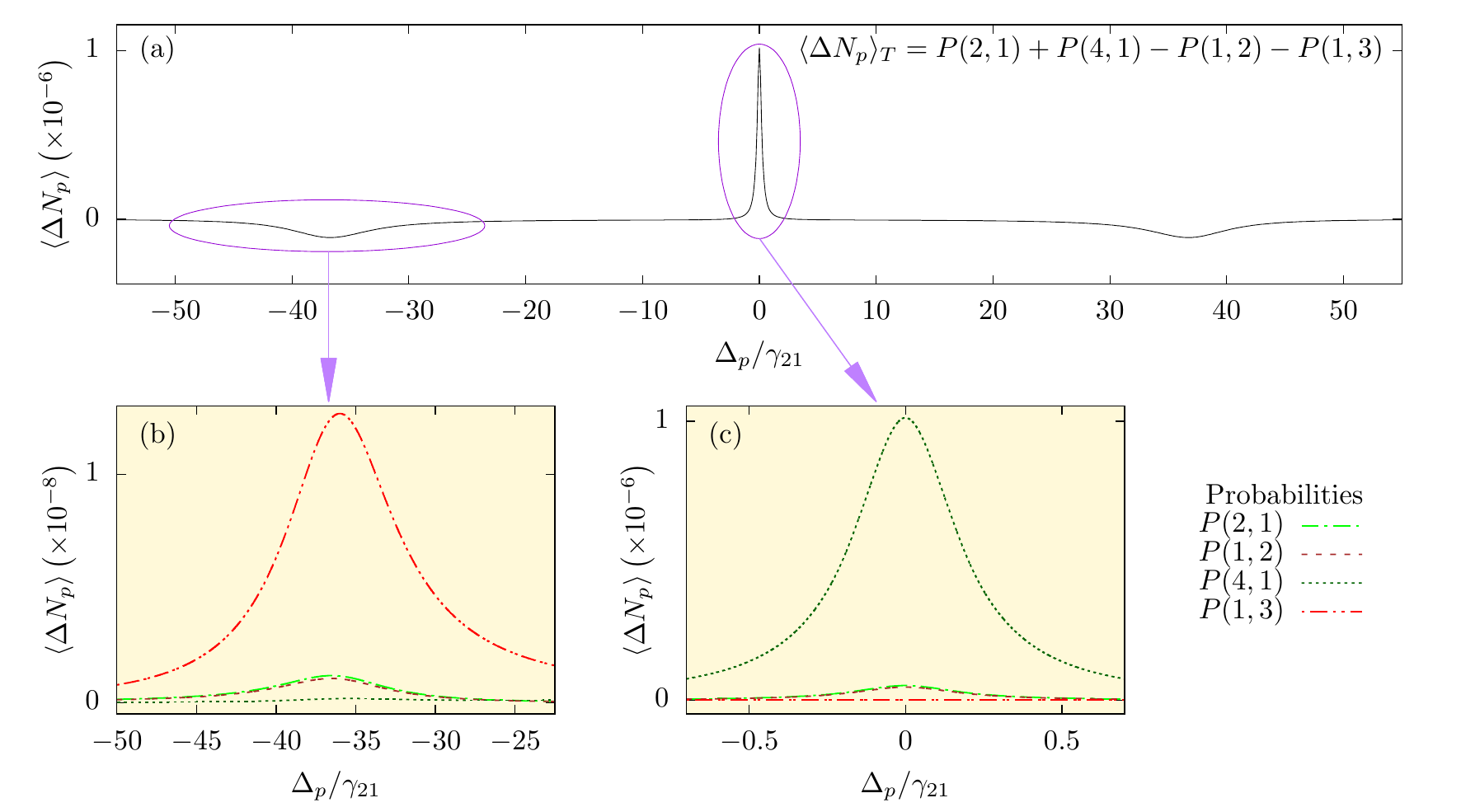}
}
\caption{(a) Total mean variation of the probe photon number per period with respect to the probe detuning $\Delta_{p}$, using $\Delta_{w}=\Delta_{s}=0$, and $\Omega_{s}=70\gamma_{21}$, which exhibits an amplification peak when the three-photon resonance condition is fulfilled, \ie $\Delta_{p}=0$, and two separated absorption peaks around $\Delta_{p}=\pm\Omega_{s}/2$. (b-c) Curves for the involved probabilities $P(i,j)$ of the corresponding regions of interest. %Note that $P(4,1)$ and $P(1,3)$ are the dominant contributions for the amplification and for the absorption regions, respectively. 
Rest of parameter values as in Fig.~\ref{fig6.3}(b).}
\label{fig6.4}
\end{figure*}
%%%%%%%%%%%%%%%%%%%%%%%%%%%%%%%%%%%%%%%%%%%%%%
%
%Because of this, the best results for probe amplification are obtained using the configuration of Fig.~\ref{fig6.3} (a) with an optimal value of $\Omega_{w}$ when the rest of parameter values are fixed. The optimal parameter for probe field amplification considering only this configuration will be discussed hereafter.
%
%%%%%%%%%%%%%%%%%%%%%%%%%%%%%%%%%%%%%%%%%%
\begin{figure*}[ht!]
\centering
{
\includegraphics[width=1\textwidth]{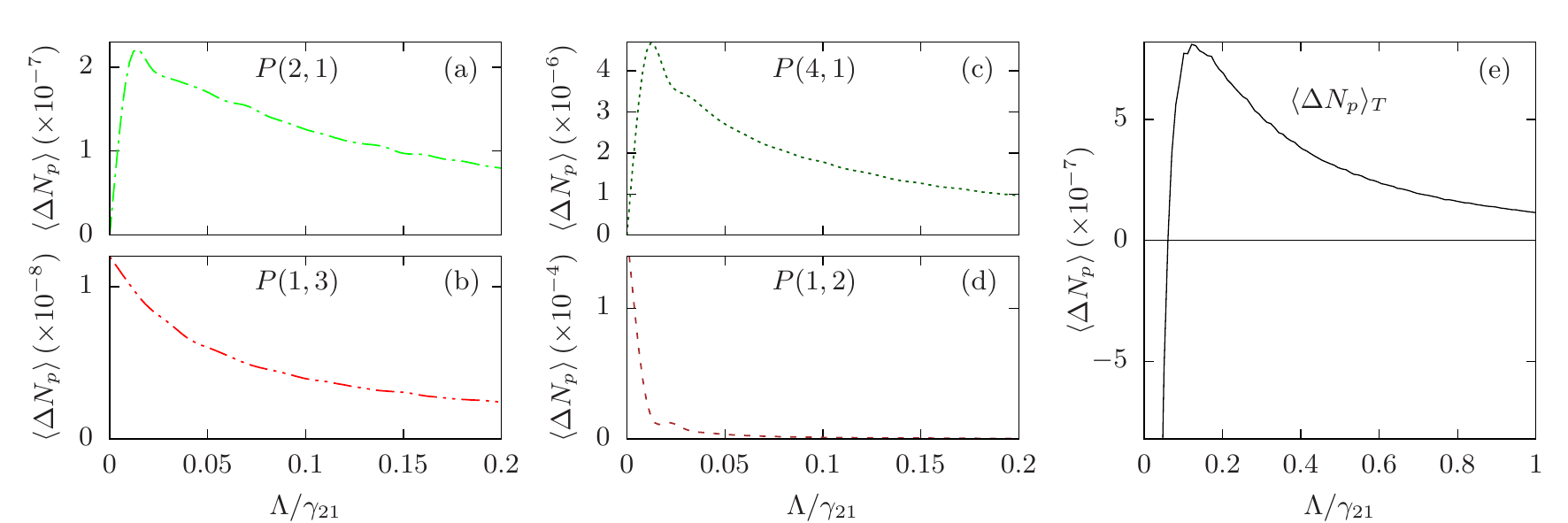}
}
\caption{Curves of (a) $P(2,1)$, (b) $P(1,3)$, (c) $P(4,1)$, and (d) $P(1,2)$ with respect to the pumping rate using $\Delta_{p}=\Delta_{w}=\Delta_{s}=0$, and the rest of the parameter values as in Fig.~\ref{fig6.3}(b), calculated using the MCFW approach. (e) Total mean variation of the probe photon number per period with respect to the pumping rate (black solid line).
}
\label{fig6.6}
\end{figure*}
%%%%%%%%%%%%%%%%%%%%%%%%%%%%%%%%%%%%%%%%%%%%%%
%

\subsection{Favourable parameter values}

Let us discuss in this Section the role of the different parameters of the scheme represented in Fig.~\ref{fig6.3}(a) in order to obtain probe field amplification.\\
\subsubsection{Emission/absorption dependence on the probe detuning}
For the most favourable configuration shown in Fig.~\ref{fig6.3}(a) it is important to highlight the fact that the two competing coherent processes, \ie period$(1,3)$ and period$(4,1)$, are not simultaneously maximised for the same detuning values of the probe laser field.
On the one hand, since $\Omega_{s}>\Omega_{w}\gg\Omega_{p}$, the absorption of the probe laser field is maximum close to $\Delta_{p}=\pm\Omega_{s}/2$.
On the other hand, period$(4,1)$ is responsible for probe stimulated emission
for values of $\Delta_{p}$ that fulfil the three-photon resonance condition, and this period does not have to compete with its opposite period, as $P(1,4)=0$.
Thus, the regions of maximum absorption and stimulated emission occur for well separated values of $\Delta_{p}$.
%However, this circumstance may change with the configuration schematised in Fig.~\ref{fig6.3} (b), as the maximum absorption and emission can be obtained for similar values of the probe detuning.

%On the other hand, period$(4,1)$ is responsible for probe stimulated emission for values of $\Delta_{p}$ that fulfil the three-photon resonance condition, \eg $\Delta_{p}=\Delta_{w}=\Delta_{s}=0$, and this period does not have to compete with its opposite period, as $P(1,4)=0$.
%
%%%%%%%%%%%%%%%%%%%%%%%%%%%%%%%%%%%%%%%%%%%%%%%%%%%%%%%%%%%%%%%%%%%%%%%%%%%%%
\begin{figure*}[ht!]
%\begin{minipage}[t]{1\textwidth}
\includegraphics[width=1\textwidth]{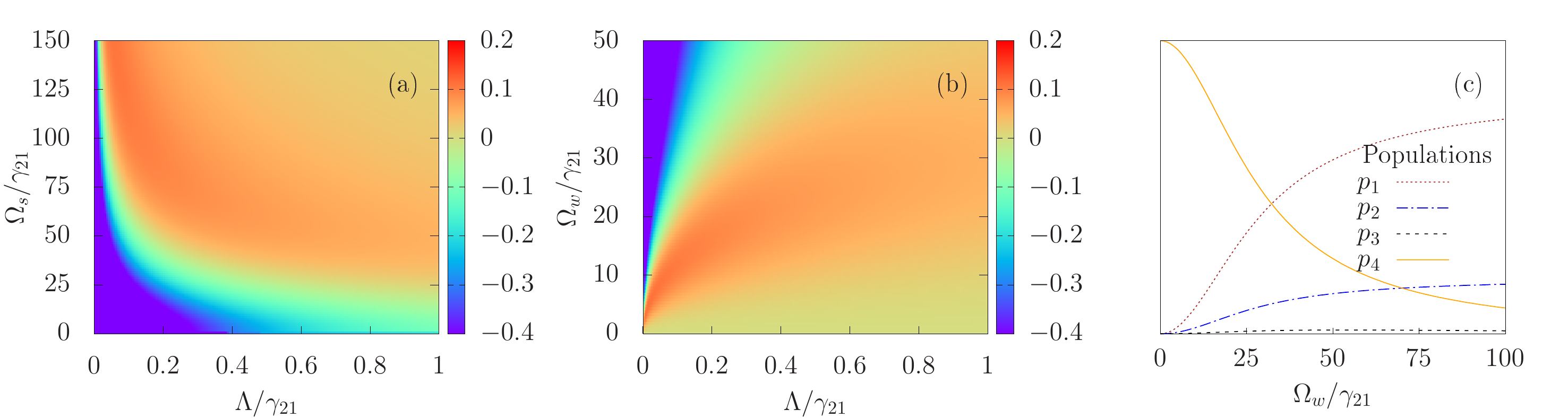}
%\end{minipage}
\caption{
Im$[\rho_{12}]/\Omega_{p}$ as a function of (a) $\Omega_{s}/\gamma_{21}$ and $\Lambda/\gamma_{21}$, using $\Omega_{w}=20\gamma_{21}$, and (b) $\Omega_{w}/\gamma_{21}$ and $\Lambda/\gamma_{21}$, using $\Omega_{s}=70\gamma_{21}$. The probe detuning is $\Delta_{p}=0$, and the rest of parameter values are the same as in Fig.~\ref{fig6.3}(b). In (c), the stationary populations $p_{i}$ of each level $\ket{i}$, with $i=1,2,3,4$, have been represented  as a function of $\Omega_{w}/\gamma_{21}$, using $\Omega_{s}=70\gamma_{21}$, and  $\Lambda=0.3\gamma_{21}$.
}
\label{fig6.7}
\end{figure*}
%%%%%%%%%%%%%%%%%%%%%%%%%%%%%%%%%%%%%%%%%%%%%%%%%%%%%%%%%%%%%%%%%%%%%%%%%%%%%
%
%Thus, the regions of maximum absorption and stimulated emission occur for well separated values of $\Delta_{p}$.
%However, this circumstance may change with the configuration schematised in Fig.~\ref{fig6.3} (b), as the maximum absorption and emission can be obtained for similar values of the probe detuning.
%
A numerical example is shown in Fig.~\ref{fig6.4} corresponding to the configuration represented in Fig.~\ref{fig6.3}(a), using $\Delta_{w}=\Delta_{s}=0$, corresponding to the experimental parameters of Ref.~\cite{Rein'15}.
The rest of parameter values are the same as the ones used in Fig.~\ref{fig6.3}(b).
Fig.~\ref{fig6.4}(a) shows the total mean variation of the probe photon number per period, $\langle\Delta N_{p}\rangle_{T}$, with respect to the probe detuning, $\Delta_{p}$, which has been obtained using \refeq{eq:Pijs} and which is identical that is obtained using the density matrix equations (DME) for ${\rm Im}\left[\rho_{12}\right]$.
In Fig.~\ref{fig6.4}(a), we observe a narrow amplification peak around $\Delta_{p}=0$ and two wide absorption peaks close to $\Delta_{p}=\pm\Omega_{s}/2=35\gamma_{21}$.
Both regions of interest are analysed in Fig.~\ref{fig6.4}(b) and Fig.~\ref{fig6.4}(c), where the curves for the probabilities of the gain probe coherent processes, $P(2,1)$ and $P(4,1)$, and for the loss probe coherent processes, $P(1,2)$ and $P(3,1)$, are shown.
Note that around the amplification peak in $\Delta_{p}=0$, the contribution of $P(1,3)$ is negligible while $P(4,1)$ is the dominant one.
As it can be seen in Fig.~\ref{fig6.4}(b) and Fig.~\ref{fig6.4}(c), two-photon loss processes, associated to P(1,3), do not take place at $\Delta_p=0$ but at $\Delta_p \sim \pm \Omega_{s}/2$ due to the AC-Stark splitting that the strong field produces on state \2.
This is consistent with the experimental results of Ref.~\cite{Rein'15} showing maximum amplification at $\Delta_{w}=\Delta_{s}=\Delta_{p}=0$.\\
%The opposite situation is observed around the absorption peaks. In Fig.~\ref{fig6.5} (a), we observe an amplification peak close to $\Delta_{p}=\Delta_{w}=25\gamma_{21}$ and two wide absorption peaks close to $\Delta_{p}=\pm\Omega_{s}/2=35\gamma_{21}$ \[\cf Fig.~\ref{fig6.3} (f)\].While the region of interest featured on Fig.~\ref{fig6.5} (b) shows nearly the same behaviour as in Fig.~\ref{fig6.4} (b), the one showed in Fig.~\ref{fig6.5} (c) presents a maximum of $P(4,1)$, which is a consequence of the three-photon resonance, accompanied by a lower maximum of $P(1,3)$.
%This effect, due to the proximity of the absorption zone, is detrimental for amplification.
%Therefore, in order to obtain large probe amplification, the configuration of Fig.~\ref{fig6.3} (a) is more preferable than that of Fig.~\ref{fig6.3} (b), as it avoids the undesired effect of the accompanying loss peak of $P(1,3)$.
%
%%%%%%%%%%%%%%%%%%%%%%%%%%%%%%%%%%%%%%%%%%
\begin{figure*}[ht!]
\centering
{
\includegraphics[width=1\textwidth]{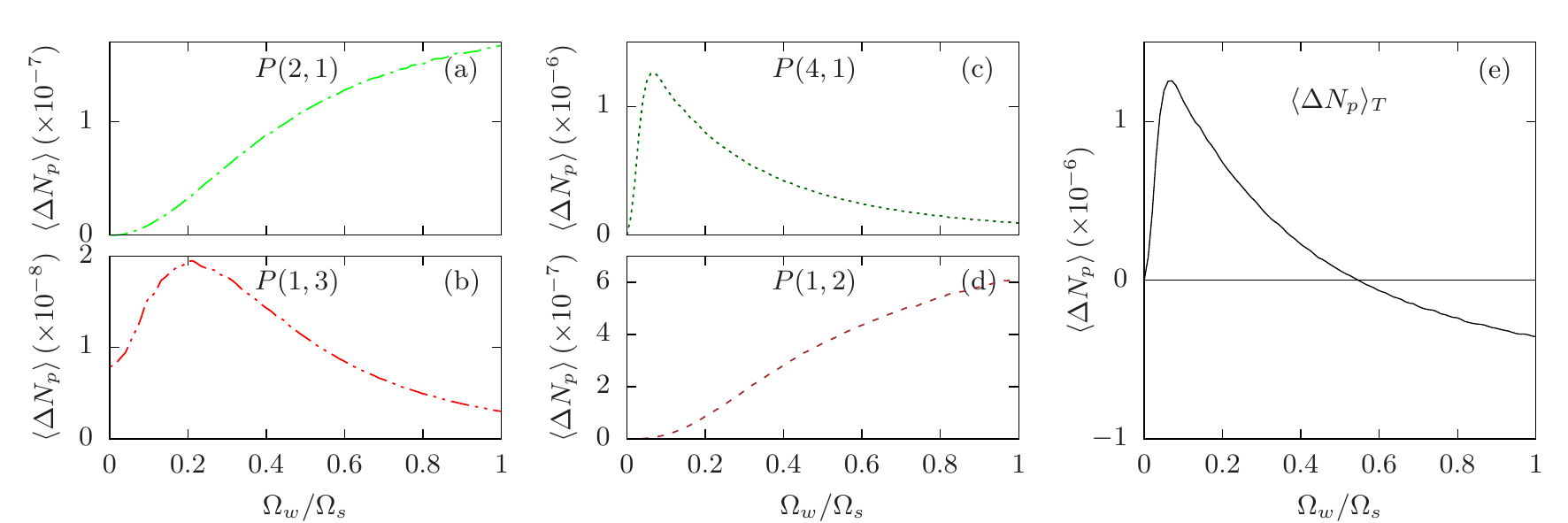}
}
\caption{Curves of (a) $P(2,1)$, (b) $P(1,3)$, (c) $P(4,1)$, and (d) $P(1,2)$ as a function of $\Omega_{w}/\Omega_{s}$ using $\Delta_{p}=\Delta_{w}=\Delta_{s}=0$, and the rest of the parameter values as in  Fig.~\ref{fig6.3}(b), calculated using the MCFW formalism. (e) Total mean variation of the probe photon number per period with respect to the ratio $\Omega_{w}/\Omega_{s}$ (black solid line).
}
\label{fig6.8}
\end{figure*}
%%%%%%%%%%%%%%%%%%%%%%%%%%%%%%%%%%%%%%%%%%%%%%
%
\subsubsection{Pumping rate}
The presence of the pumping mechanism favours probe gain processes corresponding to period$(2,1)$ and period$(4,1)$, since it allows a way
for a subsequent quantum jump that starts in \1 to take place.
Both contributions are null if there is no pumping, as can be inferred from \refeq{eq:Pijsa} and \refeq{eq:Pijsc}.
Note that, from \refeqs{eq:Pijs}, all quantities $P(i,j)\rightarrow 0$ when $\Lambda\rightarrow\infty$, since the probability of finding the system in any of the states during a coherent process in a differential time $dt$ decreases when the probability of a quantum jump that connects the states involved increases.
As a numerical example, Figs.~\ref{fig6.6}(a-e) show the dependence of the relevant probabilities $P(i,j)$ on the pumping rate for $\Omega_{s}=70\gamma_{21}$, $\Omega_{w}=20\gamma_{21}$, $\Delta_{p}=\Delta_{w}=\Delta_{s}=0$, and the rest of the parameter values as in Fig.~\ref{fig6.3}(b).
To obtain the $P(i,j)$ values from \refeq{eq:Pij}, the probabilities $P(i)$ have been calculated via the MCWF approach using $N=5000$ atoms and $dt=0.01$, in units of $\gamma_{21}^{-1}$, while the integral factors have been calculated numerically.
On the one hand, we see that $P(2,1)$ and $P(4,1)$ increase with the pumping rate from zero to a maximum value and then decrease, as shown in Figs.~\ref{fig6.6}(a,c).
On the other hand, $P(1,3)$ and $P(1,2)$ decrease monotonously from a certain value, as seen in Figs.~\ref{fig6.6}(b,d).
Fig.~\ref{fig6.6}(e) shows the total mean variation of the probe photon number per period (black solid line) obtained using \refeq{eq:meanchangeprove}, where we observe that the probe field amplification reaches a maximum for an optimal pumping value and progressively decreases to zero in the limit  $\Lambda\rightarrow\infty$.\\
\subsubsection{Strong-weak fields configuration}
For similar decay rates from level \3, $\gamma_{34}\simeq\gamma_{32}$, the condition $\Omega_{s}>\Omega_{w}$ favours the quantum jumps via $\gamma_{34}$ over those due to $\gamma_{32}$, favouring in turn the presence of the gain probe process period$(4,1)$ in the quantum trajectory. On the other hand, if $\Omega_{w}=0$, it would not be possible to initiate any coherent process in \4, so we expect that probe amplification due to this process presents a maximum for some value that fulfils $\Omega_{s}>\Omega_{w}$.
As a first numerical example,
Im$[\rho_{12}]/\Omega_{p}$ is represented as a function of $\Lambda/\gamma_{21}$ and $\Omega_{s}/\gamma_{21}$ in Fig.~\ref{fig6.7}(a), and as a function of $\Lambda/\gamma_{21}$ and $\Omega_{w}/\gamma_{21}$ in Fig.~\ref{fig6.7}(b), using $\Omega_{w}=20\gamma_{21}$ and $\Omega_{s}=70\gamma_{21}$, respectively.
The probe detuning is $\Delta_{p}=0$ and the rest of the parameter values are the same as in Fig.~\ref{fig6.3}(b). 
It can be seen from both figures that, for a given pumping rate, maximum amplification is obtained when $\Omega_{s}>\Omega_{w}$.
This effect is more important the lower the value of the pumping rate.
In addition, of $\Omega_{s}>\Omega_{w}$, the probe field amplification can be enhanced if $\gamma_{34}>\gamma_{32}$.
Note that favouring the three-photon gain process period$(4,1)$ in this way goes against the presence of the one-photon gain process period$(2,1)$, which can occur after a quantum jump via $\gamma_{32}$.
The latter, however, can be optimised by choosing an adequate value of the pumping rate, according to \refeq{eq:lambdacond1}.
In Fig.~\ref{fig6.7}(c), the stationary population $p_{i}$ of each level $\ket{i}$, with $i=1,2,3,4$, is represented as a function of the Rabi frequency of the weak field, $\Omega_{w}/\gamma_{21}$, by fixing $\Omega_{s}=70\gamma_{21}$, and $\Lambda=0.3\gamma_{21}$.
In the absence of the weak field, the entire population is accumulated in the state \4.
In the presence of the weak field, the population is distributed mainly among the states \1, \2, and \4, with no population inversion in the probe field transition, \1$\leftrightarrow$\2.
Figs.~\ref{fig6.8}(a-e) show the dependence of the relevant probabilities $P(i,j)$ on the ratio $\Omega_{w}/\Omega_{s}$, fixing $\Omega_{s}=70\gamma_{21}$, and using the same calculation process and parameter values as in Figs.~\ref{fig6.6}(a-e).
It is observed in Fig.~\ref{fig6.8}(e) that the total mean variation of the probe photon number per period (black solid line) exhibits a positive maximum for a certain value of the ratio $\Omega_{w}/\Omega_{s}$ and then decreases and even takes negative values when this  ratio increases, thus showing a transition from amplification to absorption of the probe field [\cf~Fig.~\ref{fig6.7}(b)].
This is explained as a result of the different contributions shown in Figs.~\ref{fig6.8}(a-d) according to \refeq{eq:meanchangeprove}.
Probabilities $P(2,1)$ and $P(1,2)$ have a very small value for $\Omega_{w}=0$. Both probabilities increase with the ratio $\Omega_{w}/\Omega_{s}$ in a similar way, slightly offsetting each other.
The probability $P(4,1)$ is exactly null for $\Omega_{w}=0$ and shows a maximum, as interpreted above, around $\Omega_{w}\approx0.1\Omega_{s}$.
We also see that $P(1,3)$ has a maximum around $\Omega_{w}\approx0.2\Omega_{s}$, which favours the two-photon absorption process.\\

\section{Conclusions}
\label{sec:Conclusions6}

In this work, we have used the QJ approach to study
a particular four-level atomic scheme used in neutral Hg in a recent experiment with the aim to optimise the conditions to obtain AWI for a probe field \cite{Rein'15}.
The coherent periods that contribute to the amplification and absorption of the probe field have been identified and, under certain approximations, semi-analytical expressions associated with their probabilities of occurrence along the quantum trajectory have been derived.
The most favourable atomic configurations for the probe amplification have been identified, as well as the optimal relationships between the system's parameters.
Specifically, the requirement to use a strong-weak field configuration has been discussed as well as the existence of (i) an optimal ratio between the Rabi frequencies of these two fields, and (ii) an optimal value of the incoherent pumping rate to maximise the amplification of the probe field.
In the experiment \cite{Rein'15}, the fulfilment of the three-photon resonance condition is required in order to have a Doppler-free system.
However, in the present study it has been shown that the three-photon resonance condition is required to obtain the maximum amplification of the probe field, \ie it fixes the best favourable configuration. The obtained results offer a deeper knowledge of the coherent processes involved in AWI and the relationships between the parameters of the four-level atomic scheme being considered.

\section*{ACKNOWLEDGMENTS}
\label{sec:ACKNOWLEDGMENTS}

V. Ahufinger and J. Mompart gratefully acknowledge financial support through the Spanish Ministry of Science and Innovation (MICINN) (FIS2017-86530-P), PID2020-118153GB-I00/AEI/10.13039/501100011033, and the Catalan Government (SGR2017-1646).

\bibliography{BibVC}

\end{document}